# Using fractional differentiation in astronomy

**Amelia Carolina Sparavigna[1] and Paul Milligan[2]**
1 Dipartimento di Fisica, Politecnico di Torino, C.so Duca degli Abruzzi 24, Torino, Italy
2 British Astronomical Association, Isle of Man Astronomical Society, Isle of Man.

**Abstract**
In a recent paper, published at arXiv:0910.2381, we started a discussion on the new possibilities arising from the use of fractional differential calculus in image processing. We have seen that the fractional calculation is able to enhance the quality of images, with interesting possibilities in edge detection and image restoration. Here, we want to discuss more deeply its role as a tool for the processing of astronomical images. In particular, the fractional differentiation can help produce a 'content-matter' based image from a pretty astronomical image that can be used for more research and scientific purposes, for instance to reveal faint objects galactic matter, nebulosity, more stars and planetary surface detail.

**1. Introduction**
Fractional calculus is generalizing derivative and integration of a function to non-integer order [1-3]. This generalization is a rather old problem, as demonstrated by a correspondence, which lasted several months in 1695, between Leibniz and L'Hopital [4]. Many other famous scientists of the past studied and contributed to the development of fractional calculus in the field of pure mathematics [5]. First applications of fractional calculus rose in 1920: only recently, it was applied to image processing [6], where it can be rather interesting for filtering and edge detection [7-9].
In a previous paper [10], we have discussed how the fractional calculus can provide benefit to image processing. In particular, we have seen how it works in edge detection, as proposed in Ref.7, and for enhancing the image quality. The paper ended with examples of fractional analysis of astronomical images. Here, we propose and discuss other examples to show more effectively the results that image processing of astronomical images can receive from fractional differentiation.
The objective of all astrophotography is to obtain as much detail as possible on a targeted cosmic image. To achieve this, very long exposure times are required sometimes of many hours. Long exposure astrophotography is vulnerable to various types of 'noise' such as atmospheric pollution, atmospheric moisture, light pollution, amp glow, dead pixels and stray light, all of which will appear in the final exposure. To produce a final image all of the astrophotography exposures are stacked on top of one another, thus increasing the level of image detail but at the same time increasing the level of noise. Here is where the problem arises; removing this noise also depletes the image detail and in consequence objects appear fainter and weaker in contrast. The result is a 'noise free' cleaner looking image, however less detail is revealed within the astronomical object. As previously proposed [10], fractional differentiation can help scan and examine an image to detect the faint, weaker edges and details to enhance and display them. It is then an image processing method suitable for astronomical photography.
Enhancing astronomical objects via fractional differentiation can be used to reveal a greater level of galactic matter, nebulosity, more stars and planetary surface detail. This could prove invaluable in helping to discover supernovae, comets, and very faint galaxies within astronomical images.

Before proposing examples of fractional differentiation applied to astronomical objects, let us shortly revise the evaluation of derivatives and gradient vectors in the fractional calculus.

**2. Discrete fractional derivatives for image processing**

Allowing integration and derivation of any positive real order, the fractional calculus can be considered a branch of mathematical analysis, which deals with integro-differential equations. Then, the calculus of derivatives is not straightforward as the calculus of integer order derivatives (the reader can find concise descriptions of this calculus in Refs.[11] and [12]).

Since image processing is usually working on quantized and discrete data, we discuss just the discrete implementation of fractional derivation. We have to deal with two-dimensional image maps, which are two-dimensional arrays of pixels, each pixel having three colour tones. The recorded image is a discrete signal, because it is discrete the position of the pixel in the map. Moreover, the signal is quantized, since the colour tones are ranging from 0 to 255. To define the partial derivatives suitable for calculations on the image maps, let us define the discrete fractional differentiation in the following way.

Let us suppose to have a signal $s(t)$, where $t$ can have only discrete values, $t = 1,2,...,n$ [13,14], the fractional differentiation of this signal is given by:

$$\frac{d^v s(t)}{dt^v} = s(t) + (-v)s(t-1) + \frac{(-v)(-v+1)}{2}s(t-2) + \frac{(-v)(-v+1)(-v+2)}{6}s(t-3) + ... \quad (1)$$

where $v$ is a real number. If we have a two-dimensional map $s(x,y)$, where $x,y$ can have only discrete values, that is $x = 1,2,...,n_x$ and $y = 1,2,...,n_y$, the partial derivatives are:

$$\frac{\partial^v s(x,y)}{\partial x^v} = s(x,y) + (-v)s(x-1,y) + \frac{(-v)(-v+1)}{2}s(x-2,y) +$$
$$+ \frac{(-v)(-v+1)(-v+2)}{6}s(x-3,y) + ...$$
$$\frac{\partial^v s(x,y)}{\partial y^v} = s(x,y) + (-v)s(x,y-1) + \frac{(-v)(-v+1)}{2}s(x,y-2) + \quad (2)$$
$$+ \frac{(-v)(-v+1)(-v+2)}{6}s(x,y-3) + ...$$

We can also define the fractional gradient [13]:

$$\nabla^v = \frac{\partial^v}{\partial x^v}\mathbf{u}_x + \frac{\partial^v}{\partial y^v}\mathbf{u}_y = G_x^v \mathbf{u}_x + G_y^v \mathbf{u}_y \quad (3)$$

where $\mathbf{u}_x, \mathbf{u}_y$ are the unit vectors of the two space directions. Given the two components of the gradient, we easily evaluate the magnitude $G^v = ((G_x^v)^2 + (G_y^v)^2)^{1/2}$. In the case when the fractional order parameter is $v = 1$, we have the well-know gradient.

In image processing, the magnitude of the gradient $G^v(x,y,c)$ is evaluated on the function $s(x,y)$ given by the image map $b(x,y,c)$ for each colour tone $c$. Partial derivatives contain only the first four terms in Eq.2 and 3. For each colour, we find the maximum value $G^v_{Max}(c)$ on the image map. Let us define the output map as in the following:

$$b_G(x,y,c) = 255 \left( \frac{G^v(x,y,c)}{G^v_{Max}(c)} \right)^\alpha \qquad (4)$$

where $\alpha$ is a parameter suitable to adjust the image visibility.

To understand the role of fractional differentiation and its possible use, let us consider the image in Fig.1. We observe a galaxy and stars in the background and in front of it. We evaluated the fractional gradient and obtained images (1.b)-(1.d) with different values of the fractional parameter $v$. The value of $\alpha$ is fixed at 0.5. Map (1.b), obtained for $v = 1$, which is working as the usual edge detector based on the ordinary gradient, almost removes the galaxy and puts in evidence the stars. Maps (1.c) and (1.d) are obtained with fractional values, $v = 0.7$ and $v = 0.4$, respectively. Note that the galaxy regains its visibility, when parameter $v$ approaches the zero value. In this case we have the original image, with a resulting appearance according to $\alpha$ parameter.

As in the case of Fig.4 in Ref.10, Figure 1 tells us that we are able to detect the faint stars in the image background. An interesting application could be the detection of comets or asteroids by comparing images of the same region of space may be in automatic detection tools. Let us remember that fractional differentiation has a behaviour different from that of integer derivatives and then the results we can obtain by applying the fractional gradient are different from those obtained by means of usual image processing tools, such as GIMP, for instance. These programs in fact have filtering actions based on integer order differentiation. GIMP and other tools are suitable for a further processing of the map obtained from fractional gradient evaluation, to have an enhancement of colours, brightness and contrast.

**3. Appreciating the galactic structures**

Being dependent on local tone variations, the application of a fractional gradient to an image is able to enhance the visibility of galactic structures, which are depending on the density matter variation. Let us follow the simple approach, which consists in testing the effect of the fractional gradient (3) on images of galaxies.

Image (2.a) shows the Whirlpool Galaxy. (2.b) and (2.c) are the maps obtained from the magnitude of gradient $G^v$, with $v = 1.0, \alpha = 0.3$ and for $v = 0.7, \alpha = 0.5$ respectively. The image was obtained by means of a video camera, unguided, with an exposure time of 149 video frames. Note the deep impact of fractional calculation on the galactic details visibility. Image (2.d) is obtained from (2.c) enhancing brightness and contrast with GIMP. In all the output images, we can see clearly many stars, which can be quite useful for comparison with images from other sources. Let us note that a simple increase of contrast and brightness of image (2.a) is not able to give a result with the same quality in the detection of small objects and of galactic structures. Let us stress that, being dependent on local tone variations, the evaluation of fractional gradient helps in enhancing the visibility of density matter variation.

The following example suggests another possible approach to galaxy observation. Using two maps, obtained with different fraction parameters $v$, the image difference can be created. Let us consider the image of Bode's galaxy, M81. Figure (3.a) shows M81: (3.b) and (3.c) are the maps obtained from the magnitude of gradient $G^\nu$, for $\nu = 0.3, \alpha = 0.4$ and for $\nu = 0.9, \alpha = 0.4$ respectively. In this case too, the image was obtained by means of an unguided video camera. These examples show that tools based on fractional gradient can be suitable for the automatic detection of faint bodies.

Image (3.d) is the difference between images (3.b) and (3.c). In this case, we have separated the galactic structure from stars. Note in (3.d) and in its larger reproduction, Figure 4, the visibility enhancement of the density structure. Another example is shown in Fig.5, with a processing of Andromeda Galaxy image. This image was recorded by means of a camera with a long exposure time. Images (5.b) and (5.c) are the maps obtained from the magnitude of gradient $G^\nu$, for $\nu = 0.3, \alpha = 0.5$ and for $\nu = 0.9, \alpha = 0.5$ respectively. And image (5.d) is the difference. Such a processing could prove quite useful in helping to discover supernovae.

**4. Conclusions**

There are many open questions concerning the application of fractional calculus to image processing [13,14], and these questions surely deserve further theoretical analysis. We prefer moving in the framework of a more practical approach to the problem, that is, we investigate categories of images, which are able to gain benefit from the fractional approach.

As a result of those examples here discussed, we suggest that fractional differentiation could properly enter those image-processing tools devoted to the detection of faint objects in astronomical images. For these images, removing the noise produced in image recording depletes the image detail, and, consequently, objects are fainter and weaker in contrast. Images are cleaner looking image but possess less detail. The fractional differentiation scan restores weaker edges and details and this could be an interesting source for further scientific purposes, for instance to discover supernovae, comets, and very faint galaxies.

**Figure captions**

**Fig.1**. Image (a) shows M33, a galaxy in the constellation Triangulum. Imaging was obtained by means of a Takahashi Epsilon 160 Astrograph Imaging Scope and a Canon EOS400 DSLR camera, manually guided. The exposure time was 2 hours and 30 minutes at ISO 800. Images (b), (c) and (d) are the maps obtained from the magnitude of gradient $G^v$, for $v = 1.$, $v = 0.7$ and $v = 0.4$ respectively. For the three images, we have $\alpha = 0.5$. Note how the fractional calculation has a deep impact on stars and galaxy visibility. (b) almost removes the galaxy, enhancing the visibility of stars in the background. Image (d) enhances the visibility of galaxy details.

**Fig.2**. Image (a) shows M51, the Whirlpool Galaxy in the constellation Canes Venatici. Imaging was obtained by means of a William Optics FLT110 [TEC] Imaging Scope and a GSTAR-EX CCD Video Camera, unguided. The exposure time was of 149 video frames. Images (b) and (c) are the maps obtained from the magnitude of gradient $G^v$, for $v = 1.,\alpha = 0.3$ and for $v = 0.7,\alpha = 0.5$ respectively. Note how fractional calculation has a deep impact on the galactic detail visibility. Image (d) is obtained by (c) enhancing brightness and contrast with GIMP.

**Fig.3**. Image (a) shows M81, a Bode's Galaxy, in the constellation Ursa Major. Imaging was obtained by means of a Takahashi Epsilon 160 Astrograph Imaging Scope and GSTAR-EX CCD Video camera, unguided. The exposure time was of 149 video frames. Images (b) and (c) are the maps obtained from the magnitude of gradient $G^v$, for $v = 0.3,\alpha = 0.4$ and for $v = 0.9,\alpha = 0.4$ respectively. Image (d) is the difference between (b) and (c). In this case, we have separated the galactic structure from the stars images.

**Fig.4**. The same as in Fig.3.d: note the enhanced visibility of those galactic structures, which are depending on the density matter variation.

**Fig.5**. Image (a) shows Andromeda Galaxy. Imaging was obtained by means of a Takahashi Epsilon 160 Astrograph Imaging Scope and Canon EOS400 DSLR camera, manually guided. Exposure time was 50 minutes at ISO 1600. Images (b) and (c) are the maps obtained from the magnitude of gradient $G^v$, for $v = 0.3,\alpha = 0.5$ and for $v = 0.9,\alpha = 0.5$ respectively. Image (d) is the difference between (b) and (c).

High resolution figures downloadable from http://staff.polito.it/amelia.sparavigna/galaxy/ and http://www.eyetotheuniverse.com/

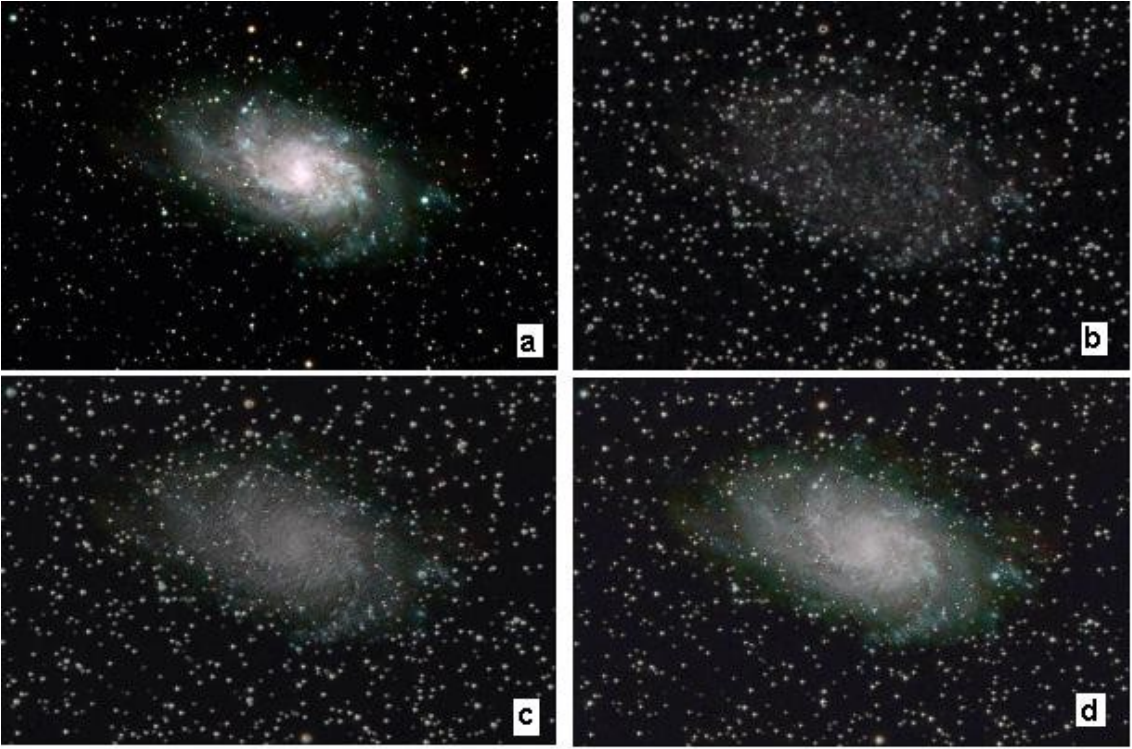

**Fig.1**

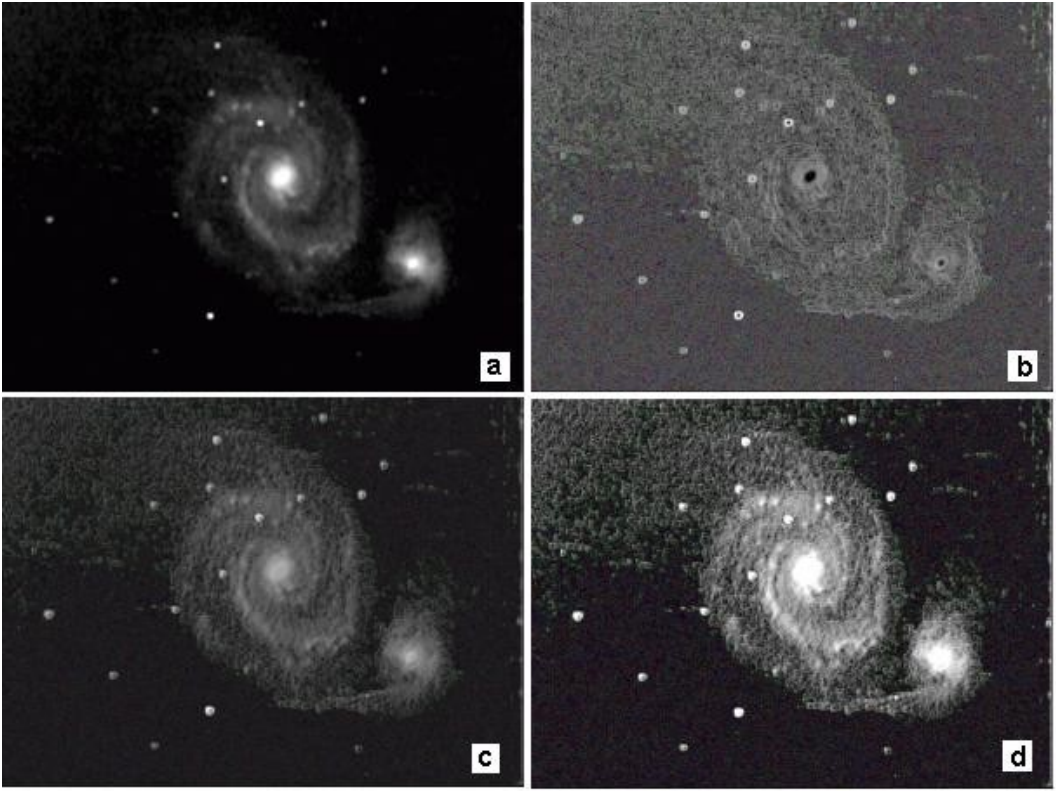

**Fig.2**

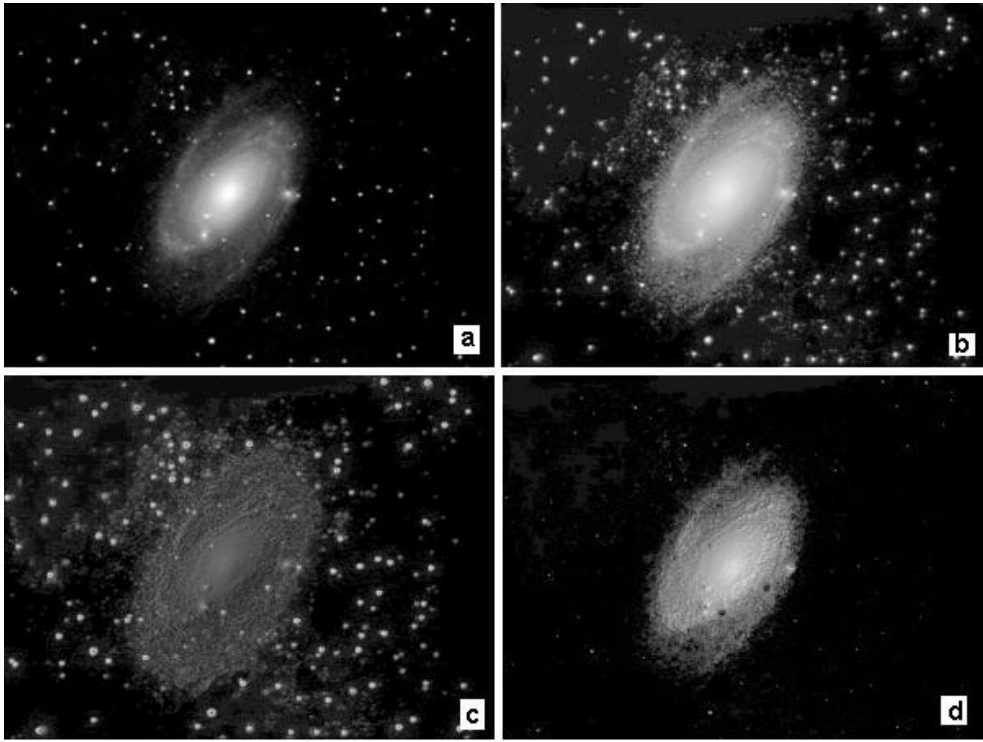

Fig.3

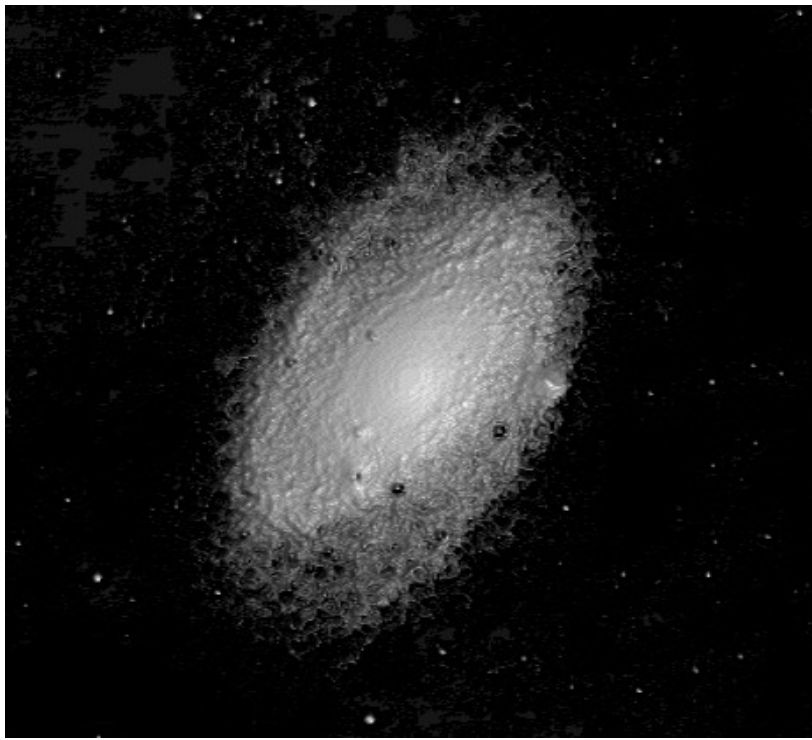

Fig.4

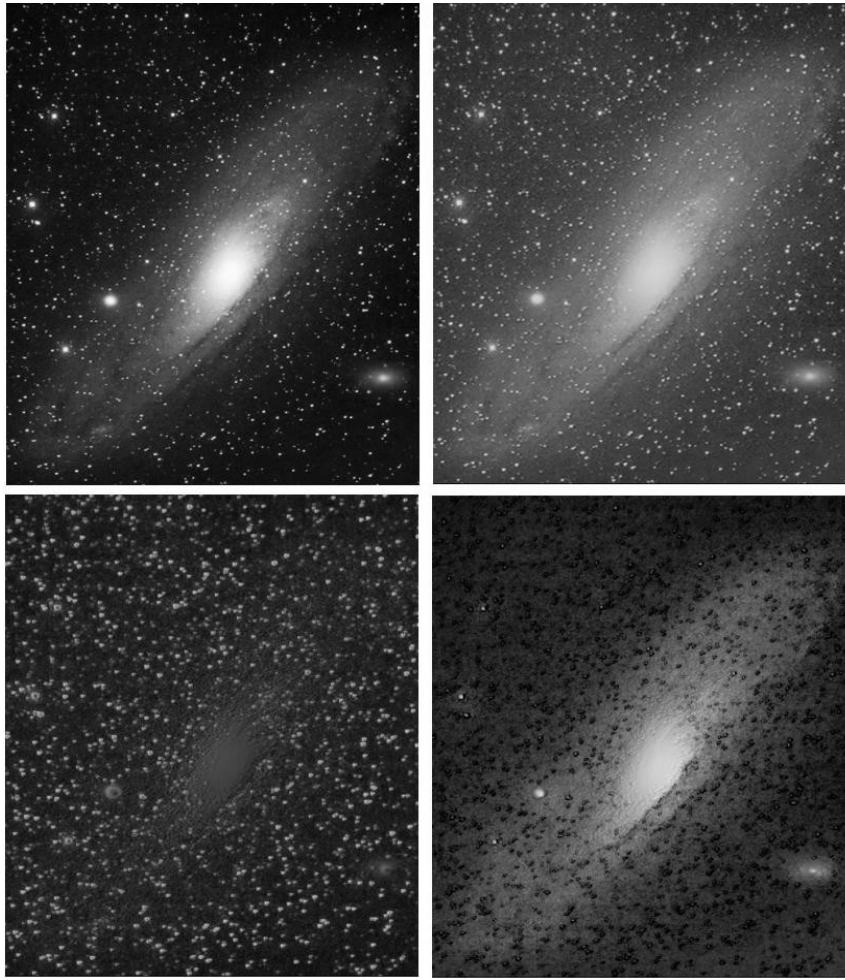

**Fig.5**